\documentclass[runningheads]{llncs}

\usepackage{graphicx}
\usepackage{cite}
\usepackage{float}
\usepackage{amsfonts}
\usepackage{amsmath}
\usepackage[colorlinks,urlcolor=blue,linkcolor=blue,citecolor=blue]{hyperref}
\usepackage[misc,geometry]{ifsym}
\usepackage{booktabs}

\begin{document}

\title{Adversarial learning of cancer tissue representations}

\author{Adalberto Claudio Quiros \inst{1} \Letter \and
Nicolas Coudray \inst{2} \and
Anna Yeaton \inst{2} \and
Wisuwat Sunhem \inst{1} \and
Roderick Murray-Smith \inst{1} \and
Aristotelis Tsirigos \inst{2} \and
Ke Yuan \inst{1} \Letter }

\authorrunning{Quiros, A.C. et al.}

\institute{University of Glasgow School of Computing Science, Scotland, UK \\
\and
New York University School of Medicine, New York, NY, USA \\
\email{a.claudio-quiros.1@research.gla.ac.uk, nicolas.coudray@nyulangone.org,} \\
\email{anna.yeaton@nyulangone.org, w.sunshem.1@research.gla.ac.uk, roderick.murray-smith@glasgow.ac.uk} \\
\email{aristotelis.tsirigos@nyulangone.org, ke.yuan@glasgow.ac.uk}}

\maketitle

\begin{abstract}
   Deep learning based analysis of histopathology images shows promise in advancing the understanding of tumor progression, tumor micro-environment, and their underpinning biological processes. So far, these approaches have focused on extracting information associated with annotations. In this work, we ask how much information can be learned from the tissue architecture itself.
    
    We present an adversarial learning model to extract feature representations of cancer tissue, without the need for manual annotations. We show that these representations are able to identify a variety of morphological characteristics across three cancer types: Breast, colon, and lung. This is supported by 1) the separation of morphologic characteristics in the latent space; 2) the ability to classify tissue type with logistic regression using latent representations, with an AUC of $0.97$ and $85\%$ accuracy, comparable to supervised deep models; 3) the ability to predict the presence of tumor in Whole Slide Images (WSIs) using multiple instance learning (MIL), achieving an AUC of $0.98$ and $94\%$ accuracy.
    
    Our results show that our model captures distinct phenotypic characteristics of real tissue samples, paving the way for further understanding of tumor progression and tumor micro-environment, and ultimately refining histopathological classification for diagnosis and treatment. \footnote{The code and pretrained models are available at: \url{https://github.com/AdalbertoCq/Adversarial-learning-of-cancer-tissue-representations}}

\end{abstract}

\section{Introduction}
    Histological images, such as hematoxylin and eosin (H\&E) stained tissue microarrays (TMAs) and whole slide images (WSIs), are an imaging technology routinely used in clinical practice, that relay information about tumor progression and tumor microenvironment to pathologists. Recently, there has been advances in relating tissue phenotypes from histological images to genomic mutations \cite{Coudray2018, Kather2020, Fu2020, Coudray2020}, molecular subtypes \cite{Woerl2020, Schmauch2020}, and prognosis \cite{Coudray2020, Fu2020}. While these previous works have contributed greatly to the field, and have highlighted the richness of information in H\&E slides, they are limited by the frequently unavailable labels required to train supervised methods. 
    
    Unsupervised learning offers the opportunity to build phenotype representations based on tissue architectures and cellular attributes, without expensive labels or representations that are only correlated with a selected predicted outcome as discriminative models. 
    
    Within unsupervised models, Generative Adversarial Networks (GANs) have been widely used in digital pathology, from nuclei segmentation \cite{Mahmood2020}, stain transformation and normalization \cite{Rana2018, Zanjani2018}, to high-quality tissue samples \cite{Krause2021}. In addition, there has been some initial work on building representations of cells \cite{Hu2019} or larger tissue patches \cite{quiros2020}.
    
    In particular, PathologyGAN \cite{quiros2020} offers a Generative Adversarial Network (GAN) that captures  tissue features and uses these characteristics to give structure to its latent space (e.g. colour, texture, spatial features of cancer and normal cells, or tissue type). The generator has representation learning properties where distinct regions of the latent space are directly related to tissue characteristics. However, it does not offer the ability to project real tissue images to its latent space, only offering latent representations for synthesized tissue. Such representations are crucial if we want to relate morphological and cellular features of tissue with genetic, molecular, or survival patient information. 

    \subsubsection{Contributions} We propose an adversarial learning model to extract feature representations of cancer tissue, we show that our tissue representations are built on morphological and cellular characteristics found on real tissue images. We present a Generative Adversarial Network with representation learning properties that includes an encoder, allowing us to project real tissue onto the model’s latent space. In addition, we show evidence that these representations capture meaningful information related to tissue, illustrating their applicability in three different scenarios:
    
    1) We visualize the representations of real tissue and show how distinct regions of the latent space enfold different characteristics, and reconstructed tissue from latent representations follow the same characteristics as the original real image; 2) We quantify how informative these representation are by training a linear classifier to predict tissue types; 3) We demonstrate the suitability of the tissue representations on a multiple instance learning (MIL) task, predicting presence of tumor on WSIs.
    
    \begin{figure}[!b]
        \centering
        \includegraphics[scale=0.15]{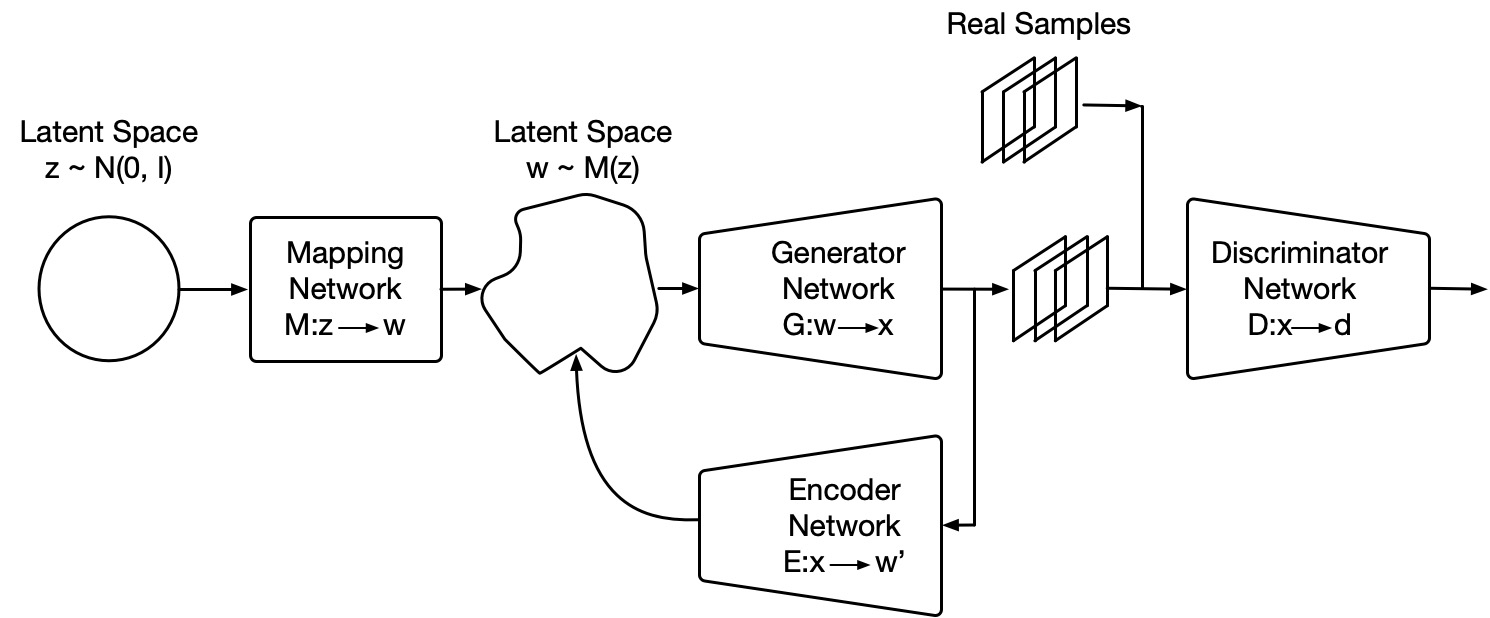}
        \caption{High level architecture of our GAN model.}
        \label{fig:pathologygan_enc_model}
    \end{figure}

\section{Method}
    
    We build upon PathologyGAN \cite{quiros2020}, and introduce an encoder $E$ that maps back generated tissue to the GAN's latent space. Effectively learning to interpret tissue morphologies and acting as the inverse of the generator $G$. After training, the encoder can be used independently to map real tissue samples to their representations in the latent space. Figure \ref{fig:pathologygan_enc_model} captures the high level architecture of our model. 
    
    The loss functions for the discriminator $L_{Dis}$ and the generator $L_{Gen}$ are defined as the Relativistic Average Discriminator \cite{Jolicoeur-Martineau2018}, where the discriminator’s goal is to estimate the probability of the real data being more realistic than the fake (Equations 1 and 2). $P_{data}$ is the distribution of real data, $G(w)$ is the distribution of synthetic data produced by the Generator $G$, $C(x)$ is the non-transformed discriminator output or critic, $W$ the transformed latent space with representation learning properties, and $P_{z} = \mathbb{N}(0,I)$ the original latent space: 
    \begin{align}
        L_{Dis}=-\mathbb{E}_{x_{r} \sim P_{data}}&\left[\log \left(\tilde{D}\left(x_{r}\right)\right)\right]-\mathbb{E}_{x_{f} \sim G(w)}\left[\log \left(1-\tilde{D}\left(x_{f}\right)\right)\right], \\
        L_{Gen}=-\mathbb{E}_{x_{f} \sim G(w)}&  \left[\log\left(\tilde{D}\left(x_{f}\right)\right)\right]-\mathbb{E}_{x_{r} \sim P_{data}}\left[\log \left(1-\tilde{D}\left(x_{r}\right)\right)\right], \\
        \quad \quad \tilde{D}\left(x_{r}\right) &= \text{sigmoid}\left(C\left(x_{r}\right)-\mathbb{E}_{x_{f} \sim G(w)} C\left(x_{f}\right)\right)), \nonumber \\ 
        \quad \quad \tilde{D}\left(x_{f}\right) &= \text{sigmoid}\left(C\left(x_{f}\right)-\mathbb{E}_{x_{r} \sim P_{data}} C\left(x_{r}\right)\right), \nonumber \\
        \quad \quad w &= M(z),\ z \sim P_{z}. \nonumber
        \label{eqn:disc_loss}
    \end{align} 
    
    We use the mean square error between latent vectors $w$ and their reconstruction from generated images $w'=E(G(w))$ as the  encoder loss function, $L_{Enc}$ (Equation \ref{eqn:enc_loss}):  
    \begin{equation}
        L_{Enc} = \mathbb{E}_{z \sim P_{z}}\left[ \frac{1}{n} \sum_{i=1}^{n}(w_{i} - w'_{i})^{2}\right] \textit{ where }
        w'=E(G(w)),\ w = M(z).
         \label{eqn:enc_loss}
    \end{equation}
    
    Although the encoder $E$ is simultaneously trained with the GAN model, we can separate the model training into two parts: The mapping network $M$, generator $G$, and discriminator $D$ that are trained as a GAN, and the encoder $E$, which is trained to project back the generated cancer tissue images onto the latent space. In practice, the encoder $E$ learns with the Generator $G$. We trained our encoder based on the assumption that the generator is successful in reproducing cancer tissue. Therefore the encoder will learn to project real tissue images if it is able to do so with generated ones. Based on this logic, we use only generated images to train the encoder. 
    
    Additionally, the encoder is only updated when the generator is not trained with style mixing regularization \cite{Karras2019}. Style mixing regularization is a technique that promotes disentanglement where the generator $G$ generates an image from two different latent vectors $w_{1}$ and $w_{2}$, these vector are feed at different layers of the generator $G$. However, it becomes impractical to train the encoder in these steps because these images have no clear assignation in the latent space $W$. Therefore, The style mixing regularization is only preformed $50$\% of times in the generator training, so our encoder is updated every two steps per the generator.
    
    \subsubsection{Datasets} 
    \label{datasets}
        We trained our model with three different cancer types: Breast, colon, and lung. We provide a detailed description on these datasets and how they were built in the Appendix \ref{Appendix:Datasets}.
        
        The breast H\&E cancer dataset is composed by the Netherlands Cancer Institute (NKI, Netherlands) and Vancouver General Hospital (VGH, Canada) cohorts \cite{Beck2011} with 248 and 328 patients, each patient with associated Tissue Micro-Arrays (TMAs). Each TMA is divided into $224\times224$ tiles and labeled subject to density of cancer cells in the tile using CRImage \cite{Yuan2012}, we use $9$ different classes with class $8$ accounting for tiles with the largest count of cancer cells. This dataset is composed by a training set of $249$K tiles and $460$ patients, and a test set of $13$K tiles and $116$ patients, with no overlapping patients between sets. 
        
        The colorectal H\&E cancer dataset from National Center for Tumor diseases (NCT, Germany) \cite{kather_2018} provides tissue images of $224\times224$ resolution with an associated type of tissue label: Adipose, background, debris, lymphocytes, mucus, smooth muscle, normal colon mucosa, cancer-associated stroma, and colorectal adenocarcinoma epithelium (tumor). The dataset is divided into a training set of $100$K tissue tiles and $86$ patients, and a test set of $7$K tissue tiles and $50$ patients, there is no overlapping patients between train and test sets. Finally, in order to compare our model to supervised methods on tissue type classification, we combine the classes stroma and smooth muscle into a class 'simple stroma' as in reference publications \cite{Raczkowski2019, Kather2016}.
        
        The lung H\&E cancer dataset contains samples with adenocarcinoma (LUAD), squamous cell carcinoma (LUSC), and normal tissue, composed by $1807$ Whole Slide Images (WSIs) of $1184$ patients from the Cancer Genome Atlas (TCGA). Each WSI is labeled as tumor and non-tumor depending on the presence of lung cancer in the tissue and divided into $224\times224$ tiles. In addition, we split the dataset into a training set of $916$K tissue tiles and $666$ patients, and a test set of $569$K tissue tiles and $518$ patients, with no overlapping patients between both sets. 
        
\section{Results}
    Since our tissue representations are built in an unsupervised manner and from tissue features alone, we studied how meaningful real tissue representations are. We show results quantifying the tissue morphology and cellular information contained in the representations, and provide examples of how they can be exploited: Latent space visualization and tissue reconstruction, tissue type classification, and tumor prediction in a multiple instance learning setting.
    
    \subsection{Visualizations of tissue representations and reconstructions}
        
        We first aim to test how interpretable the latent representations of real tissue are, and if they capture enough information to recreate tissue with the same characteristics. We do so by visualizing the latent representations along with the corresponding real images and by reconstructing images from real tissue projections.  
          
        We used the breast cancer and colorectal cancer datasets for these results. As covered in Section \ref{datasets}, the breast tissue patches have an associated label with the density of cancer cells in the image and the colorectal tissue patches have an associated label with the corresponding tissue type. In both cases, we used the training set to train our GAN model and later used the encoder $E$ to project images from test sets, obtaining representations of real tissue samples. 
        
\begin{figure}[!b]
    		\centering
    		\minipage{0.5\textwidth}
    		  \includegraphics[scale=0.12]{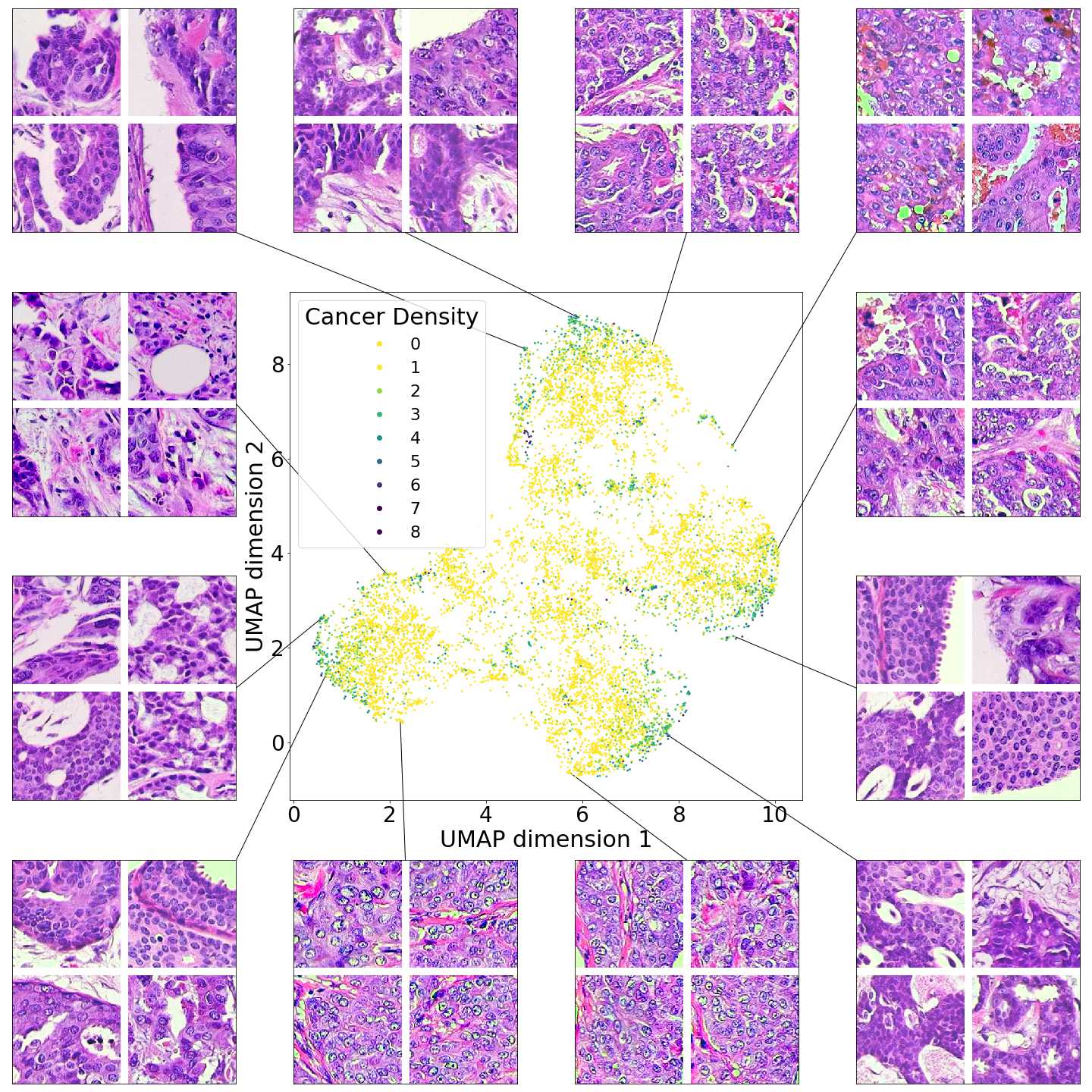}
    		\endminipage\hfill
    		\minipage{0.5\textwidth}
    		  \includegraphics[scale=0.12]{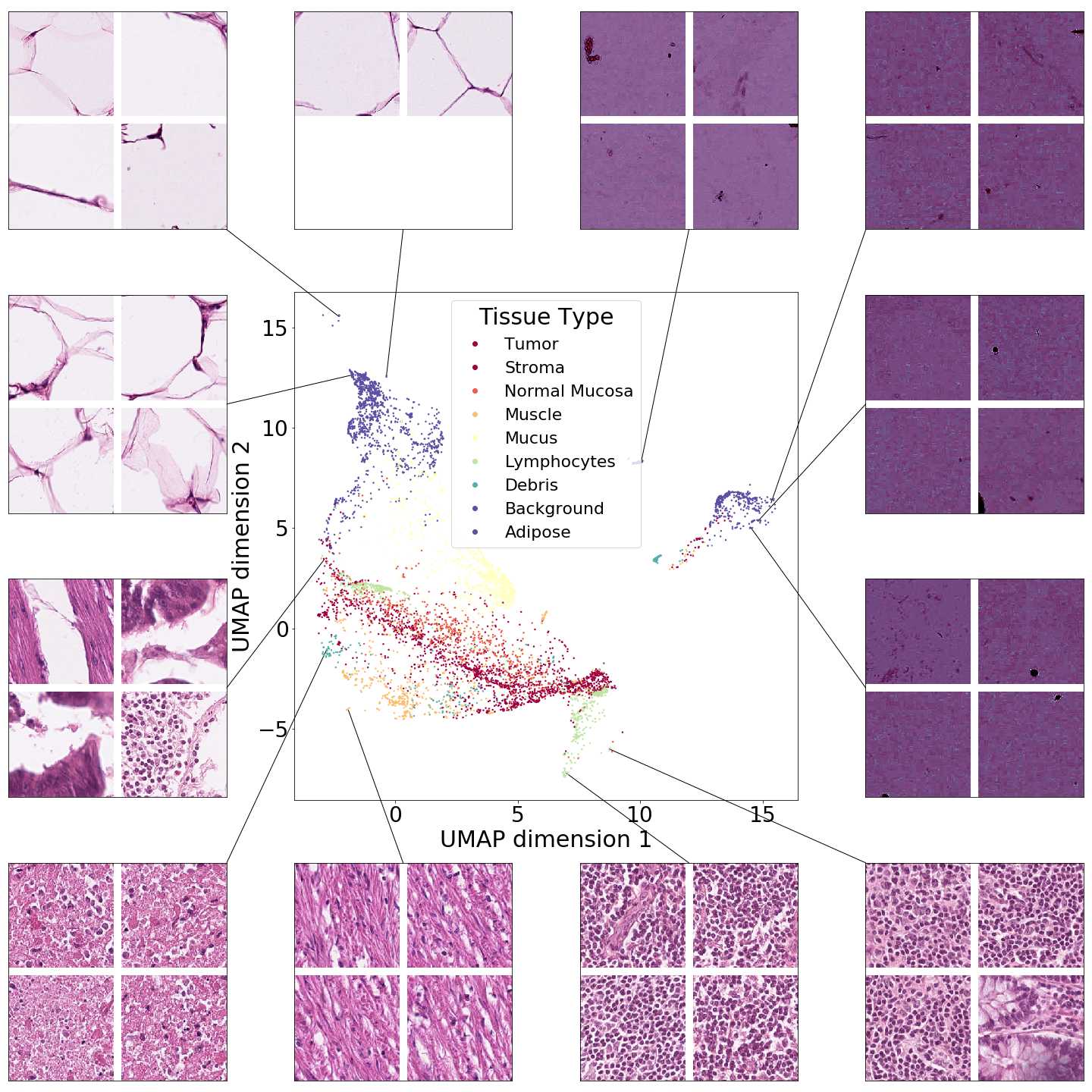}
    		\endminipage
            \caption{Uniform Manifold Approximation and Projection (UMAP) vectors of PathologyGAN's latent representations, we present breast cancer tissue from NKI and VGH (left image) and colorectal cancer tissue from NCT (right image). Breast cancer tissue images are labeled using cancer cell counts, class $8$ accounting for the largest cell number. Colorectal cancer tissue images are labeled based on their tissue type. In both cases, we observe that real tissue projected to the latent space retain cellular information and tissue morphology.}
    		 \label{fig:latent_space}
    	\end{figure}

        In Figure \ref{fig:latent_space} we used UMAP \cite{McInnes2018} to reduce the dimensionality of latent representations from $200$ to $2$ and plot each latent vector with its associated real tissue image and label. In the breast cancer case, tissue with the highest density of cancer cells is distributed outwards while no presence of them concentrates inwards. Colorectal cancer samples are located in different regions of the space depending of their tissue types. In both examples, tissue characteristics of real images, whether if it is tissue type or cancer cell density, determines the location of the representation in the latent space. 
        
\begin{figure}[!b]
            \centering
            \includegraphics[scale=0.22]{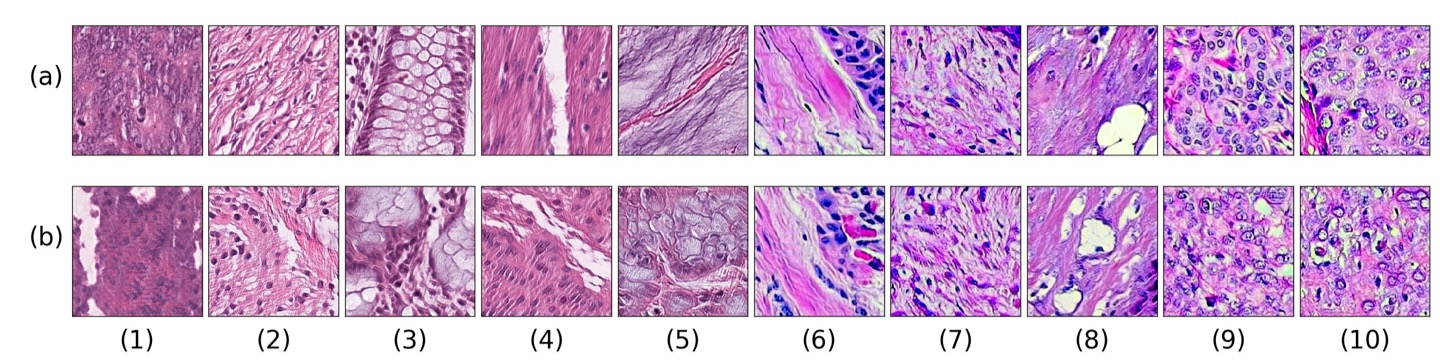}
            \caption{Real tissue images and their reconstructions. We take real tissue images and map them to the latent space with our encoder, then we use the generator with the latent vector representations to generate the image reconstructions. (a) correspond to the real tissue images $X_{real}$ and (b) to the reconstructions $X_{recon}=G(E(X_{real}))$, the images are paired in columns. We present samples of colorectal cancer tissue from NCT (1-5) and breast cancer tissue from VGH and NKI (5-10). We show different examples of tumor(1,9,10), stromal (2,8), normal mucosa (3), muscle (3), and lymphocytes (6,7), the reconstructions follow the real image attributes.}
            \label{fig:real_recon}
        \end{figure}

        In addition, we wanted to compare real images with the  synthetic images  reconstructed by the generator at the specific position in the latent space where each real image was projected, $X_{recon}=G(E(X_{real}))$. Figure \ref{fig:real_recon} shows real images and their associated reconstructions, we used latent representations of real images (a) to synthesize their reconstructions (b) with the GAN's generator. We provide examples of the colorectal cancer tissue types (1-5) and different cancer cell densities in breast cancer tissue (6-10). We can see that the reconstructions follow the characteristics of the real tissue images. 
        
        We conclude that tissue representations are not only interpretable but also hold the relevant information needed to synthesize tissue with the same characteristics. We provide further samples of latent space visualizations in the Appendix \ref{Appendix:latent_visual}
        
    \subsection{Tissue type classification over latent representations}
        In this task, we aim to quantify how informative latent representation are, verifying that the model is able to learn tissue patterns that define tissue type. We train a linear classifier to predict tissue type over latent representations. 
        
        We used the colorectal cancer dataset with each tissue sample and its associated tissue type. We trained our GAN with the training set and again used the Encoder $E$ to project the training and test sets onto the GAN's latent space, obtaining representations for each tissue image. Consecutively, we used the latent representations for the training set to train a logistic regression (one-versus-rest) and evaluate performance using test set projections. 
        
        Table \ref{table:logistic} shows the performance of the logistic regression, reaching an average AUC (one-vs-rest) of $0.976$ and multi-class accuracy of $85.43$\% just from latent representations. 
        Table \ref{table:model_comparison} provides a comparison between other existing supervised methods, such as Raczkowski et al. \cite{Raczkowski2019} where a Bayesian deep neural network is trained for the purpose of tissue classification achieving an AUC of $0.992$ and $92.44$\% accuracy, in addition, Kather et al. \cite{Kather2016} provides another performance reference using an RNF-SVM with an AUC of $0.995$ and $92.44$\% accuracy. We provide further details on ROC and confusion matrices in the Appendix \ref{Appendix:roc}
        
        Phenotype representations from our model provide robust information such that a linear classifier is able to predict tissue types with high accuracy. 
        Despite a small loss of performance when compared to the best supervised method, our fully unsupervised representations could be an effective solution when extensive and precise manual annotations cannot be obtained.
        
        \begin{table}[!t]
            \centering
            \begin{tabular}{llllllll}
                \hline
                \hline
                \multicolumn{8}{c}{\bfseries Total AUC: $\mathbf{0.976}$}  \\
                \hline
                \bfseries Tumor & \bfseries Simple Stroma & \bfseries Mucosa &  \bfseries Mucus & \bfseries Lymph. & \bfseries Debris & \bfseries Back. & \bfseries Adipose \\
                $0.974$ & $0.929$ & $0.964$ & $0.997$ & $0.994$ & $0.959$ & $1.0$ & $0.998$  \\
                \hline
                \hline
                \multicolumn{8}{c}{\bfseries Total Accuracy: $\mathbf{85.43}$\%}  \\
                \hline
                \bfseries Tumor & \bfseries Simple Stroma & \bfseries Mucosa & \bfseries Mucus & \bfseries Lymph. & \bfseries Debris & \bfseries Back. & \bfseries Adipose \\
                $89$\% & $71$\% & $68$\% & $91$\% & $83$\% & $63$\% & $100$\% & $96$\%  \\
                \hline
            \end{tabular}
            \caption{Logistic regression Accuracy and AUC on tissue type classification. We used the colorectal cancer tissue images from NCT to train a classifier over latent representations. A logistic regression (one-vs-rest) is able to reliably find information in tissue representations to predict tissue types.}
            \label{table:logistic}
        \end{table}
        
        \setlength{\tabcolsep}{12pt}
        \begin{table}[!t]
            \centering
            \begin{tabular}{l|ll}
                \hline
                \hline
                \bfseries Model & \bfseries AUC & \bfseries Accuracy \\
                \hline
                \hline
                Ours & $0.976$ & $85.43$\%  \\
                \hline
                Bayesian DNN \cite{Raczkowski2019} &  $\mathbf{0.995}$ & \bfseries $\mathbf{99.2}$\% \\
                \hline
                RBF-SVM \cite{Kather2016} & $0.976$ & $87.4$\% \\
                \hline
            \end{tabular}
            \caption{Performance comparison on tissue type classification between existing methods and our tissue latent representations. These results reflect baseline performance of supervised deep learning and non-deep learning approaches. Our representations on logistic regression without any transformation or projection, and the fact that they are comparable with supervised performance, demonstrate the applicability and information hold in them.}
            \label{table:model_comparison}
        \end{table}

    \subsection{Multiple Instance Learning on latent representations}
        Finally, we tested the reliability of the tissue phenotype representations in a weakly supervised setting. We used tissue representations in a multiple instance learning (MIL) task for tumor presence prediction in Whole Slide Images (WSIs), where each WSI has an average, minimum, and maximum of $974$, $175$, and $14$K, tiles, respectively. 
        
        We used the lung cancer dataset where each WSI has an associated label tumor or normal subject to presence of tumor tissue. We divided each WSI into $224\times224$ patches to train our GAN model and obtain tissue representations, and we later used all tissue representations of the WSI. 
        
        In the case of the MIL problem, we have a bag of instances $X = \{x_{1}, ..., x_{n}\}$ with an individual label $Y \in \{0,1\}$. We further assume that each instance of the bag has an associated label $y_{i} \in \{0,1\}$ to which we have no access:
        $$Y=\left\{\begin{array}{ll}
            0, & \text { iff } \sum_{k} y_{k}=0 \\
            1, & \text { otherwise }
        \end{array}\right.$$
        
        \begin{figure}[!b]
    		\centering
    		\includegraphics[scale=0.19]{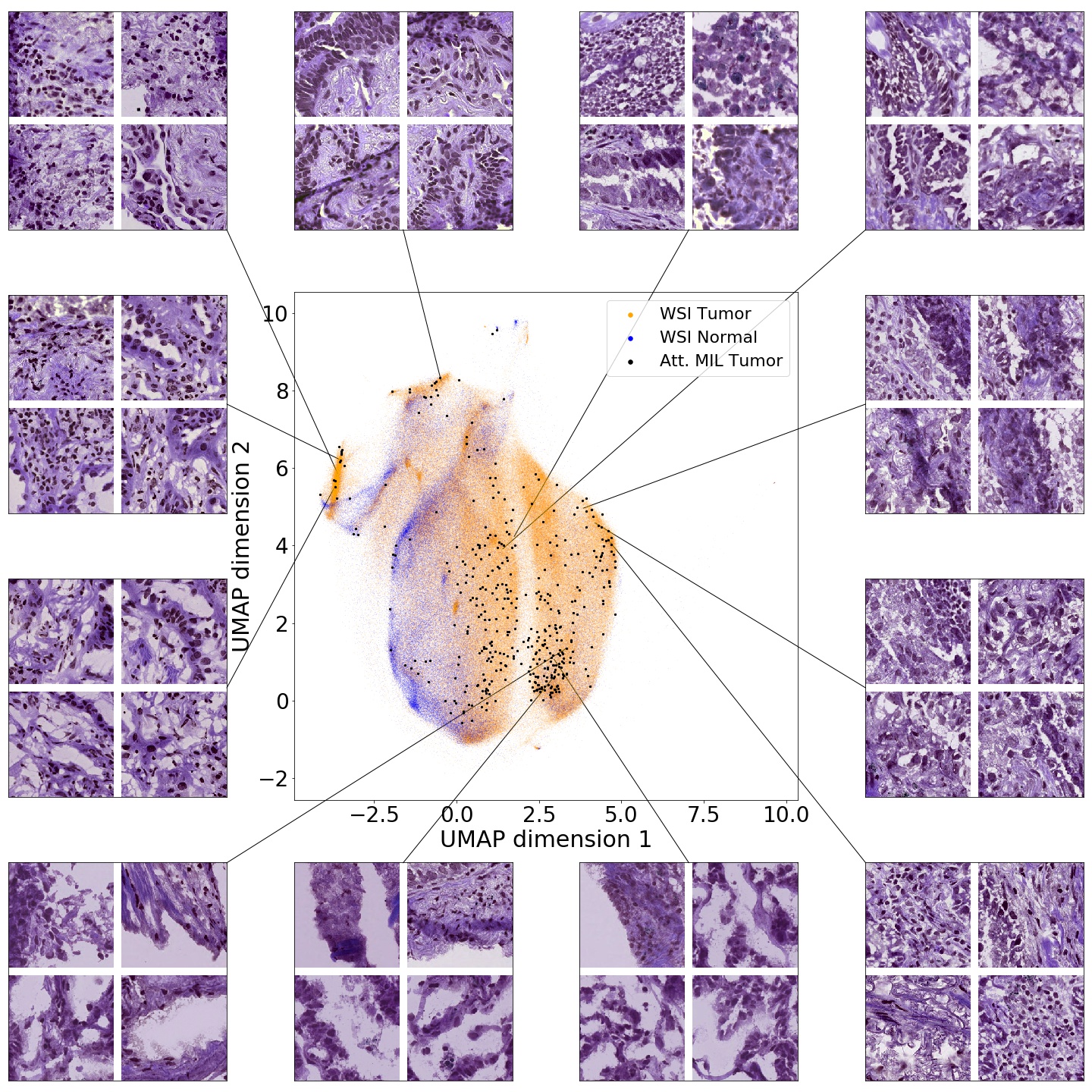}
            \caption{Uniform Manifold Approximation and Projection (UMAP) vectors of lung cancer tissue representations. We labeled each patch of the WSI with the corresponding label subject to presence of tumor in the WSI, and highlight images and representations where the attention-based deep MIL focuses to predict the outcome. We can see that the MIL framework emphasizes on regions where there is only the presence of tumor patches, relying on the information contained in the latent representations.}
    		 \label{fig:latent_space_frozen_MIL}
    	\end{figure}
        
        In our case, we have tissue tiles that we translated into tissue representations $W = \{w_{1}, ..., w_{n}\}$, and use them to determine the presence of lung tumor $Y=1$ in the WSI. We used the attention-based deep MIL \cite{ilse2018} as it assigns a weight to each instance of the bag $x_{k}$, allowing us to measure which representations are relevant for the prediction. 
    
        The attention-based deep MIL over latent representations achieves an AUC of $0.98$ and accuracy of $94$\%. These results are comparable to Coudray et al. \cite{Coudray2018}, where an Inception-V3 network was tested on the same dataset and purpose achieving an AUC of $0.993$ and accuracy of $97.5$\%.
        
        Figure \ref{fig:latent_space_frozen_MIL} shows UMAP reductions of tissue patches representations for the test set, we labeled each patch representation with the WSI label, tumor or normal depending on the presence of tumor. In addition we highlighted images and representations of the top $0.1$\% most weighted representations for the tumor outcome prediction. We can see that the MIL model focuses on regions of the latent space where there is no presence of normal tissue. 
        
        Given the accurate prediction of lung tumor presence in the WSIs and the focus of the MIL framework on representations solely present in tumor tissue, we conclude that the phenotype representations are reliable enough to capture tissue characteristics such as tumor tissue at scale.
        
\section{Conclusion}
    We presented an adversarial learning model that builds phenotype representations of cancer tissue. Our model overcomes previous limitations \cite{quiros2020} by introducing an Encoder, enabling a direct mapping between real tissue images and phenotype representations. We have illustrated the applicability of its representations in three different scenarios, providing evidence of the morphological and cellular information enfold on the real tissue representations. Furthermore, we found that these latent representations learn such stable and relevant features of histopathology images that they can be used as interpretable visualizations, as feature space of a linear classifier, or as input in a weakly supervised setting such as MIL.
    
    We envision our model to be used as a tool to characterize phenotype patterns. These learned features of histopathology images can be used to find associations between tissue and matched clinical and biological data such as genomic, transcriptomic, or survival information, contributing to a further characterize tumor micro-environment and improved patient stratification. 

\subsubsection*{Acknowledgements}

We will like to acknowledge funding support from University of Glasgow on A.C.Q scholarship, K.Y from EPSRC grant EP/R018634/1, and R.M-S. from EPSRC grants EP/T00097X/1 and EP/R018634/1. This work has used computing resources at the NYU School of Medicine High Performance Computing Facility.
    
\bibliographystyle{splncs04}
\bibliography{paper2230}

\newpage

\appendix

\section{Datasets}
\label{Appendix:Datasets}
    We provide here further details on the datasets and their processing:
    \begin{enumerate}
        \item The breast H\&E cancer dataset is composed by the Netherlands Cancer Institute (NKI) and Vancouver General Hospital (VGH) cohorts with $248$ and $328$ patients, each of them with associated Tissue Micro-Arrays (TMAs). The original TMAs have a resolution of $1128\times720$ pixels, and we split each TMA into tiles of $224\times224$, allowing them to overlap by $50$\%. We also perform data augmentation on these images, a rotation of $90^{\circ}$ and $180^{\circ}$, and vertical and horizontal inversion. We filter out images in which the tissue covers less than $70$\% of the area. In addition, we label each tile based on the density of cancer cells in the image by using CRImage \cite{Yuan2012}, a SVM classifier that quantifies the number of cancer cells, the number of other types of cells (such as stromal or lymphocytes), and the ratio of tumorous cells per area. We use nine different classes with class $8$ accounting for tiles with the largest count of cancer cells. This dataset is composed by a training set of $249K$ tiles and $460$ patients, and a test set of $13K$ tiles and $116$ patients, with no overlapping patients between sets. 
        
        \item The colorectal H\&E cancer dataset from National Center for Tumor diseases (NCT, Germany) \cite{kather_2018} provides tissue images of $224\times224$ resolution with an associated type of tissue label: adipose, background, debris, lymphocytes, mucus, smooth muscle, normal colon mucosa, cancer-associated stroma, and colorectal adenocarcinoma epithelium (tumor). All tissue images are provided with stain normalization already applied \cite{Macenko2009}. The dataset is divided into a training set of $100K$ tissue patches and $86$ patients, and a test set of $7K$ tissue patches and $50$ patients, there is no overlapping patients between train and test sets. We use this dataset to visualize latent representations and reconstructions of the real tissue images, and to test the performance of a linear classifier over those same latent representations. Finally, in order to compare our model to supervised methods, we combine the classes stroma and smooth muscle into a class 'simple stroma' as in reference publications \cite{Raczkowski2019, Kather2016}.
        
        \item The lung H\&E cancer dataset contains samples with adenocarcinoma (LUAD), squamous cell carcinoma (LUSC), and normal tissue, composed by $1807$ Whole Slide Images (WSIs) of $1184$ patients from the Cancer Genome Atlas (TCGA). We make use of the pipeline provided in \cite{Coudray2018},  diving each WSI into patches of $224\times224$ and filtering out images with less than 50\% tissue in total area and apply stain normalization \cite{Reinhard2001}. In addition, we label each slide as tumor and non-tumor depending on the presence of lung cancer in the tissue. Finally, we split the dataset into a training set of $916K$ tissue patches and $666$ patients, and a test set of $569K$ tissue patches and $518$ patients, with no overlapping patients between both sets. We use this dataset to apply multiple instance learning (MIL) over latent representations, testing the performance to predict the presence of tumor in the WSI.
    \end{enumerate}
    \newpage
    
\section{Visualization of tissue representations}
\label{Appendix:latent_visual}
    In this section we provide additional examples of visualizations of breast and colorectal cancer tissue representations.
    \begin{figure}[H]
		\centering
        \includegraphics[scale=0.115]{./images/latent_space/vgh_nki_norm/UMAP_latent_space_test_radius_1p01.jpg}
        \includegraphics[scale=0.115]{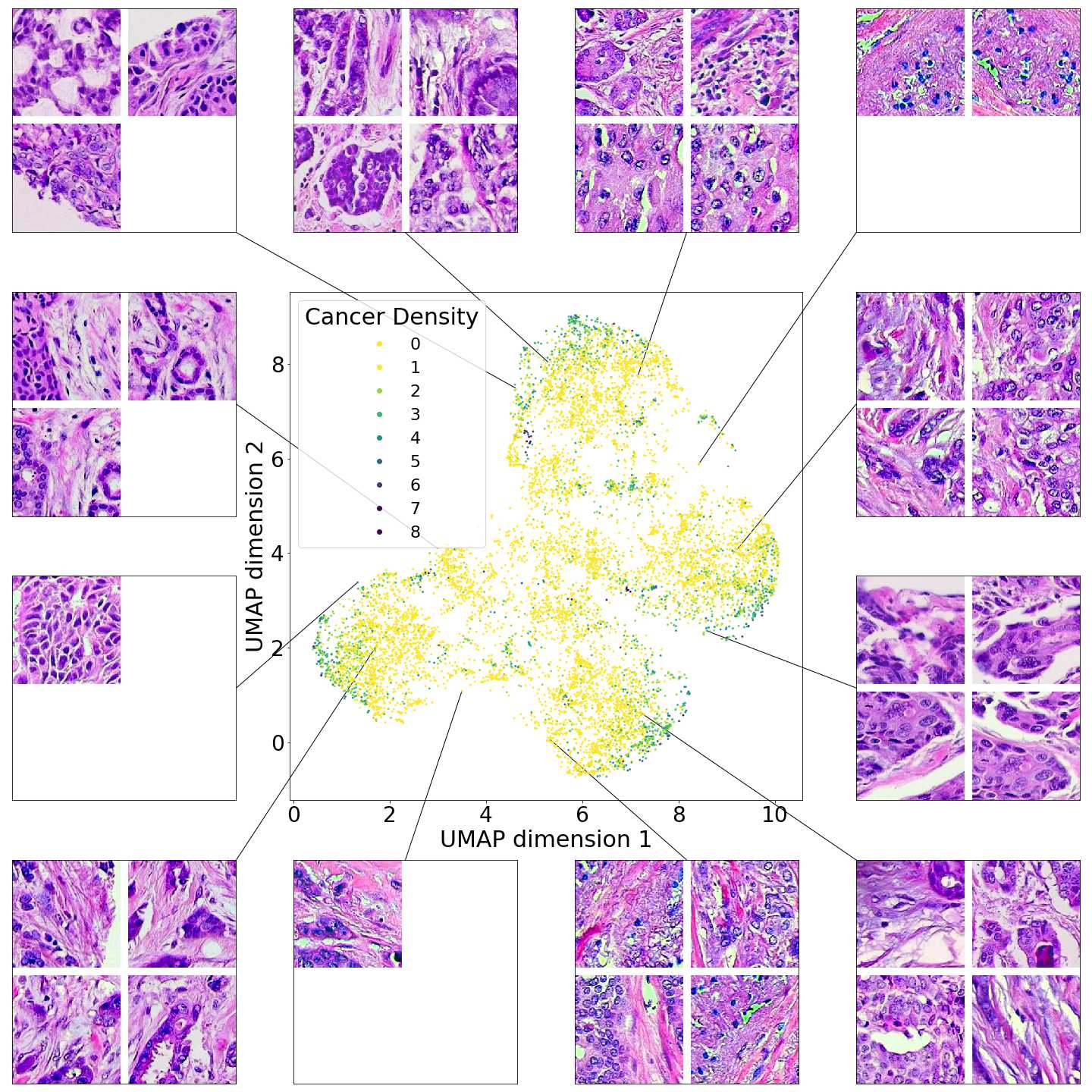}
        \includegraphics[scale=0.115]{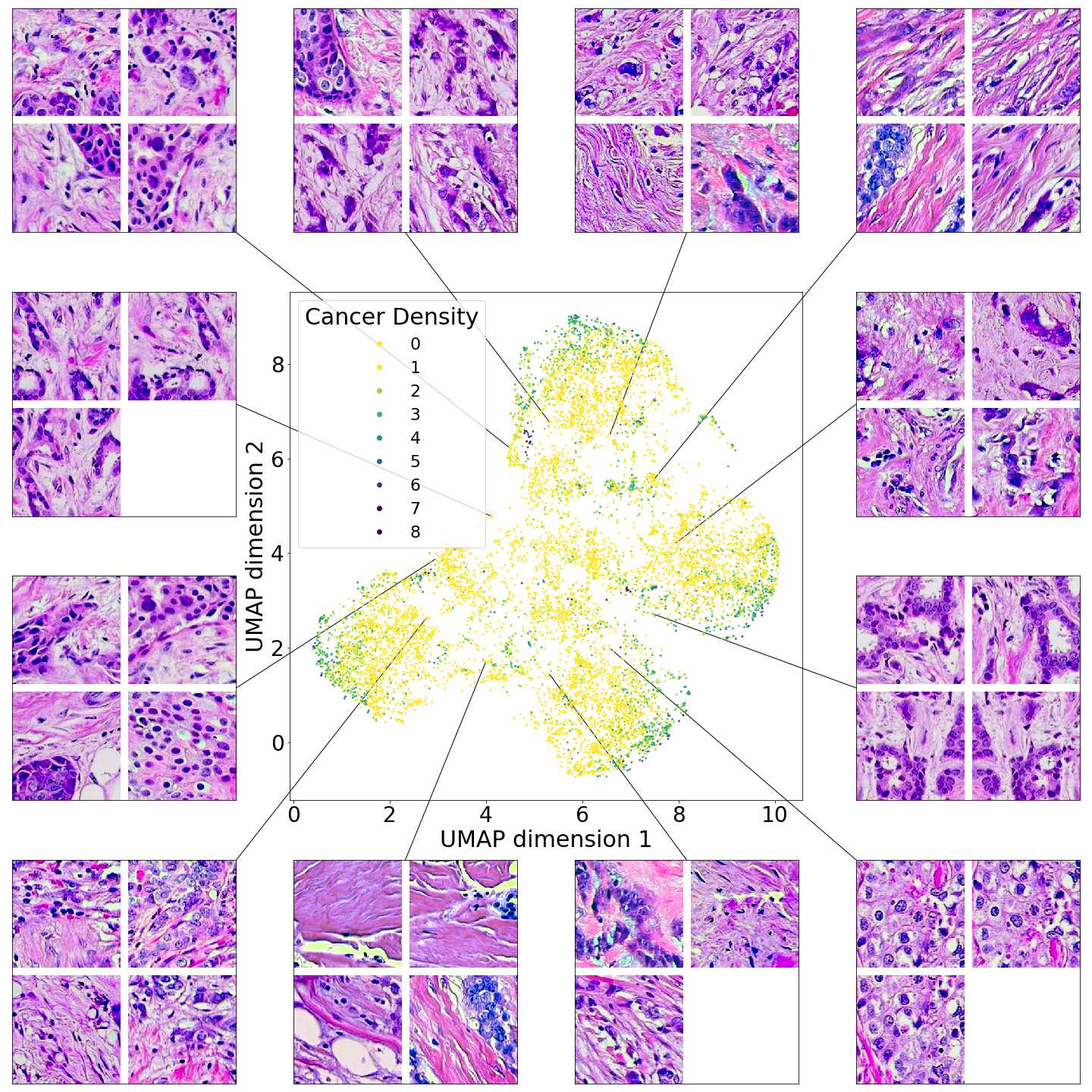}
        \includegraphics[scale=0.115]{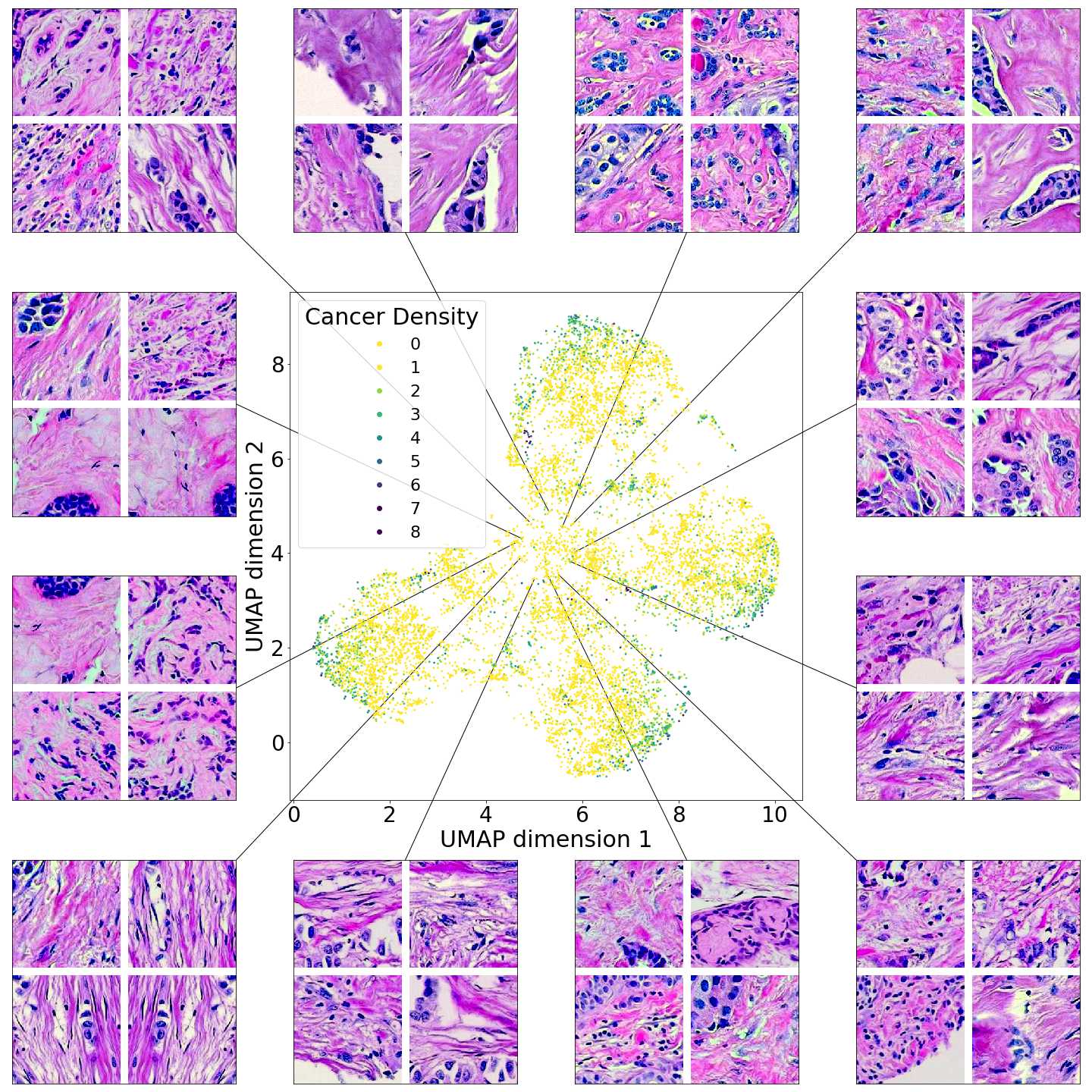}
		 \caption{Uniform Manifold Approximation and Projection (UMAP) vectors of tissue representations. We present breast cancer tissue from NKI and VGH, where images are labels based on cancer cell counts, class $8$ accounting for the largest.}
		 \label{fig:appendix_latent_space_breast}
	\end{figure}    
	
	\newpage
	
	\begin{figure}[H]
		\centering
        \includegraphics[scale=0.115]{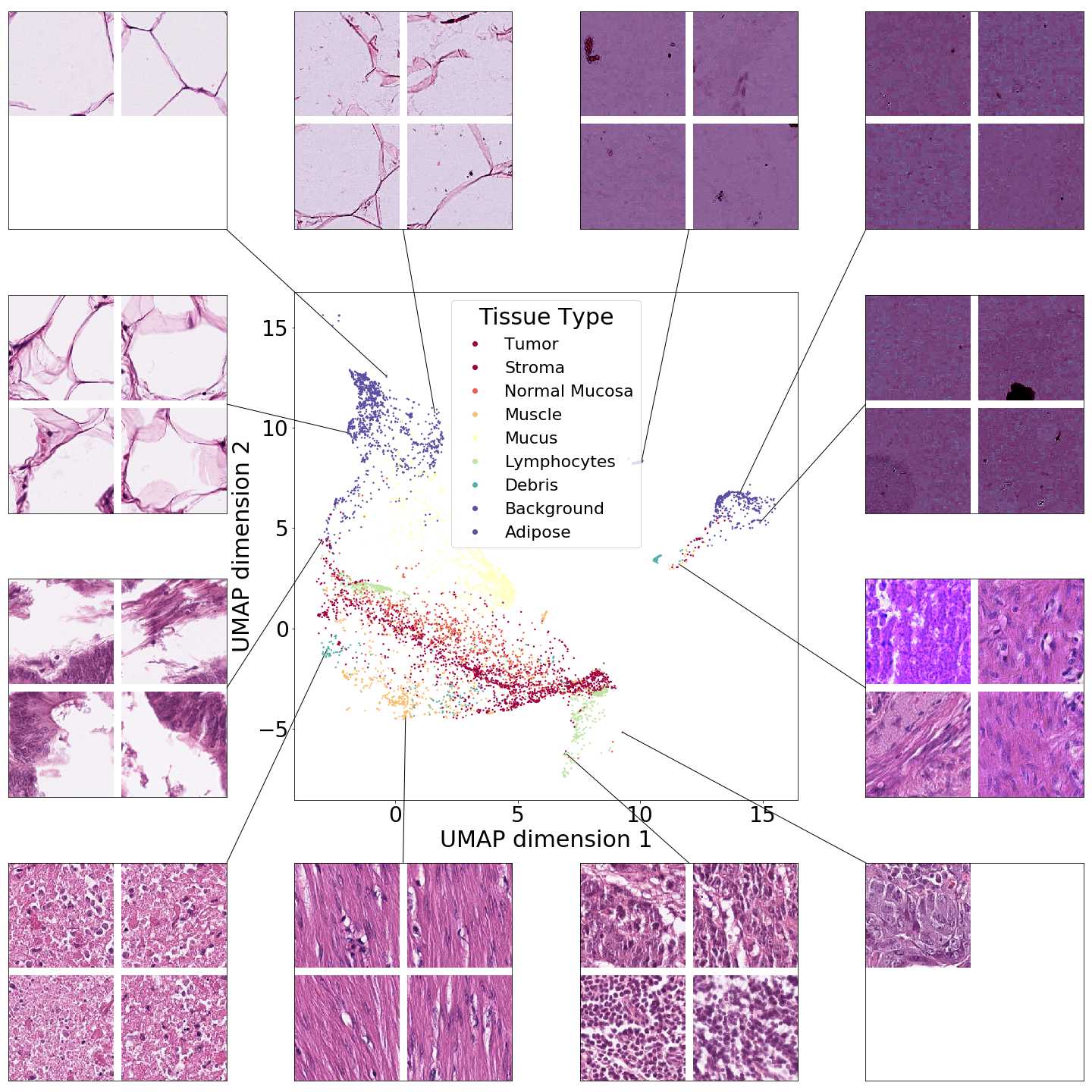}
        \includegraphics[scale=0.115]{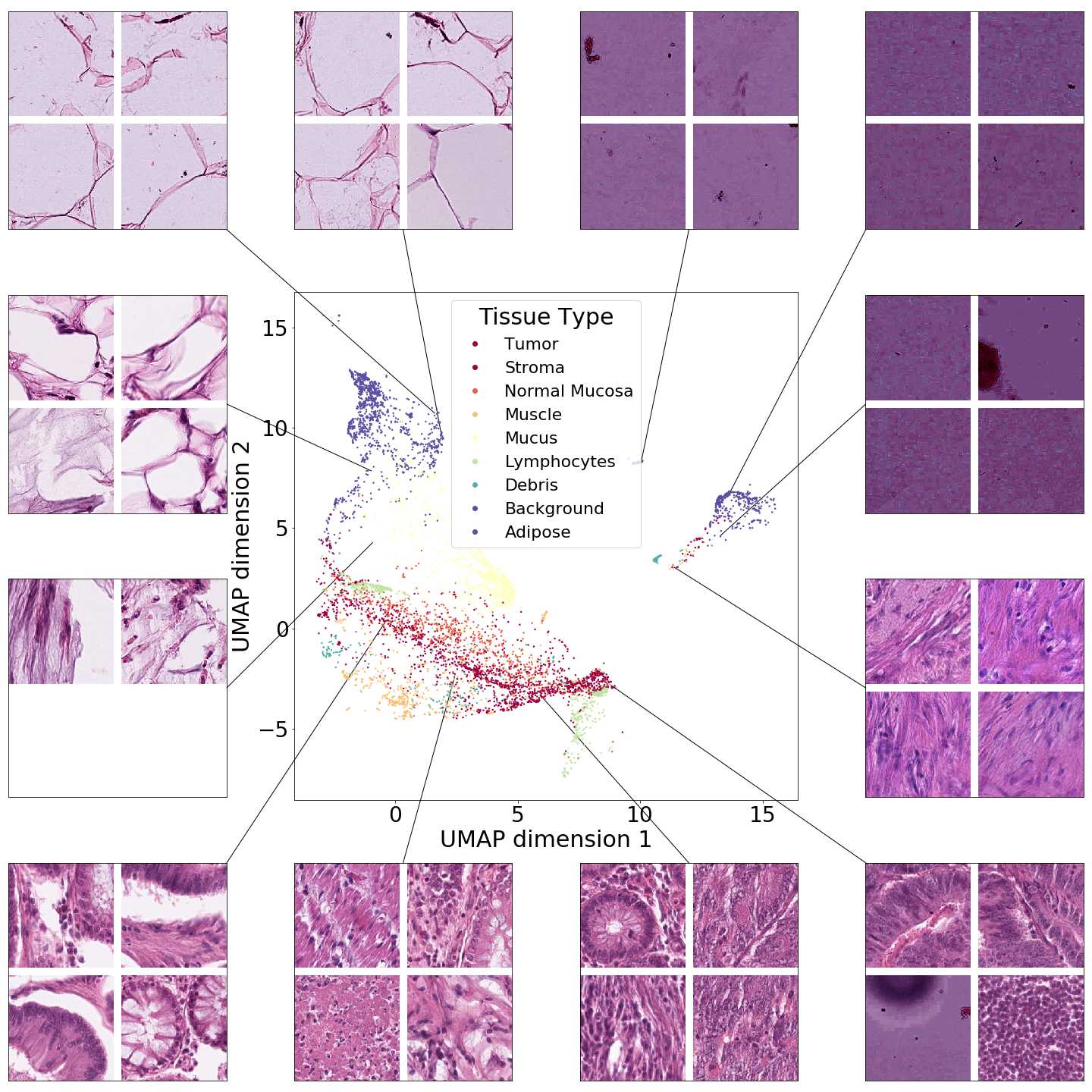}
        \includegraphics[scale=0.115]{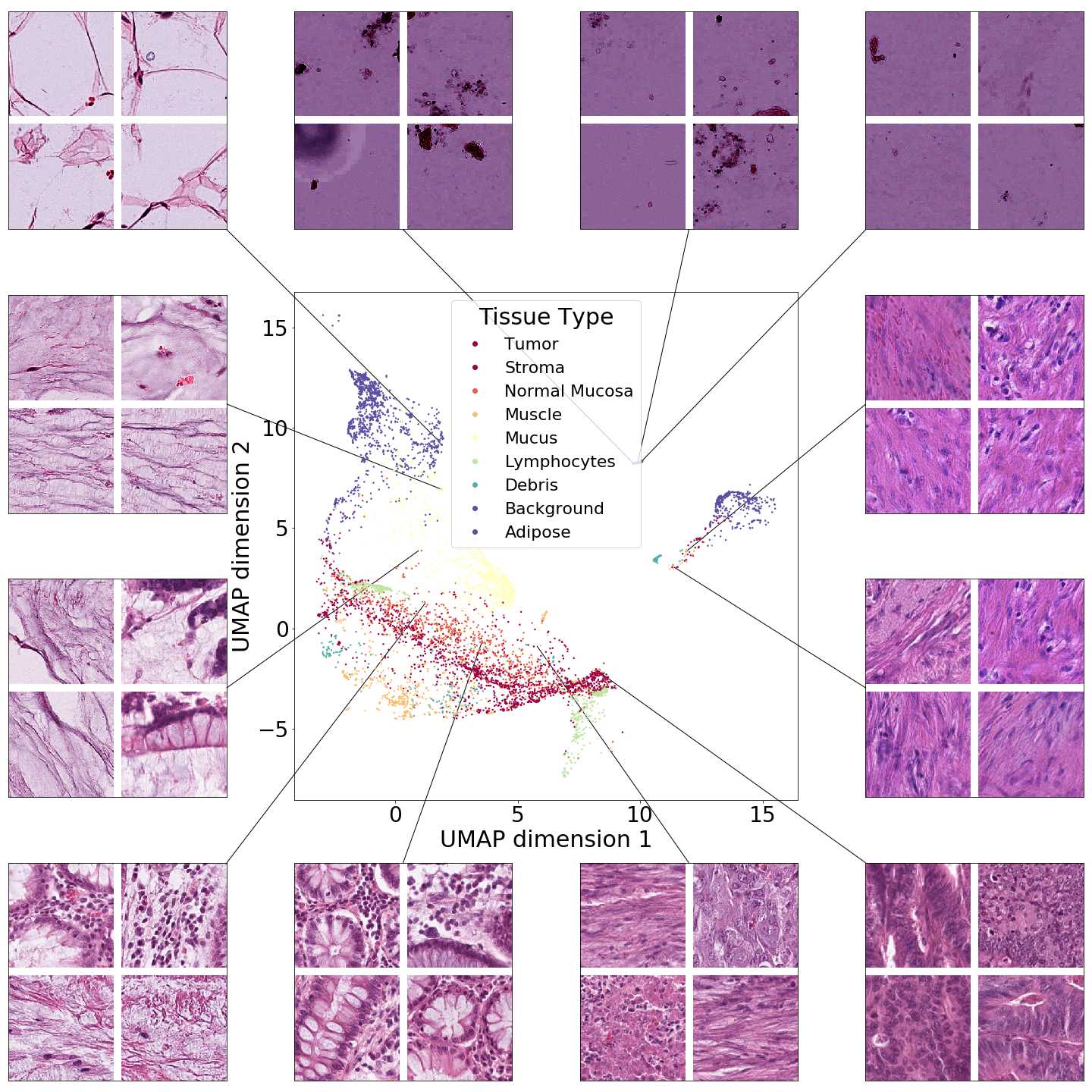}
        \includegraphics[scale=0.115]{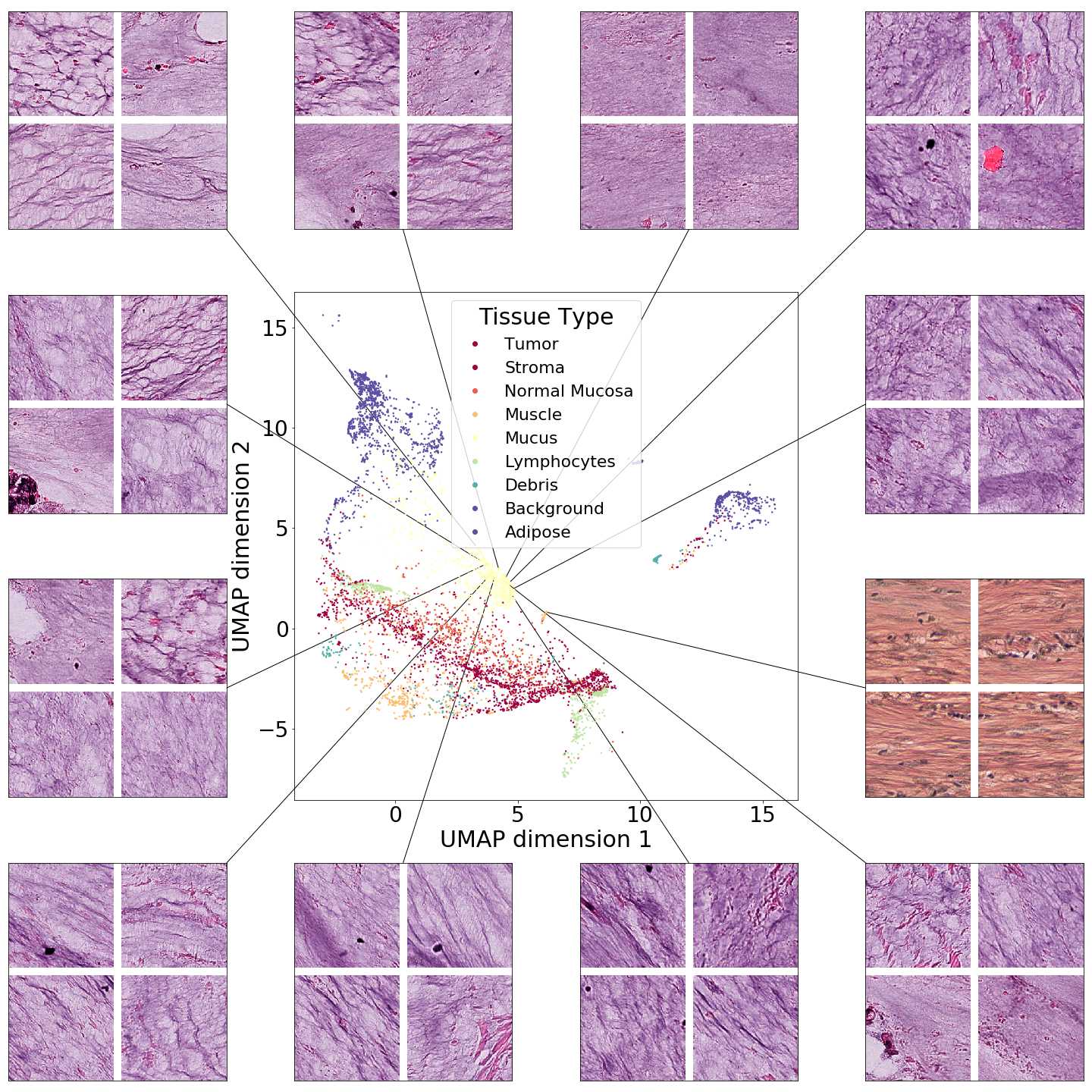}
		\caption{Uniform Manifold Approximation and Projection (UMAP) vectors of tissue representations. We present colorectal cancer tissue from NCT, where images are labeled based on the tissue type. We observe that real tissue with different morphology gets assigned distinct regions of the latent space.}
		 \label{fig:appendix_latent_space_crc}
	\end{figure}    
	
\section{Tissue Reconstructions}
\label{Appendix:reconstructions}
    \begin{figure}[H]
        \centering
        \includegraphics[scale=0.24]{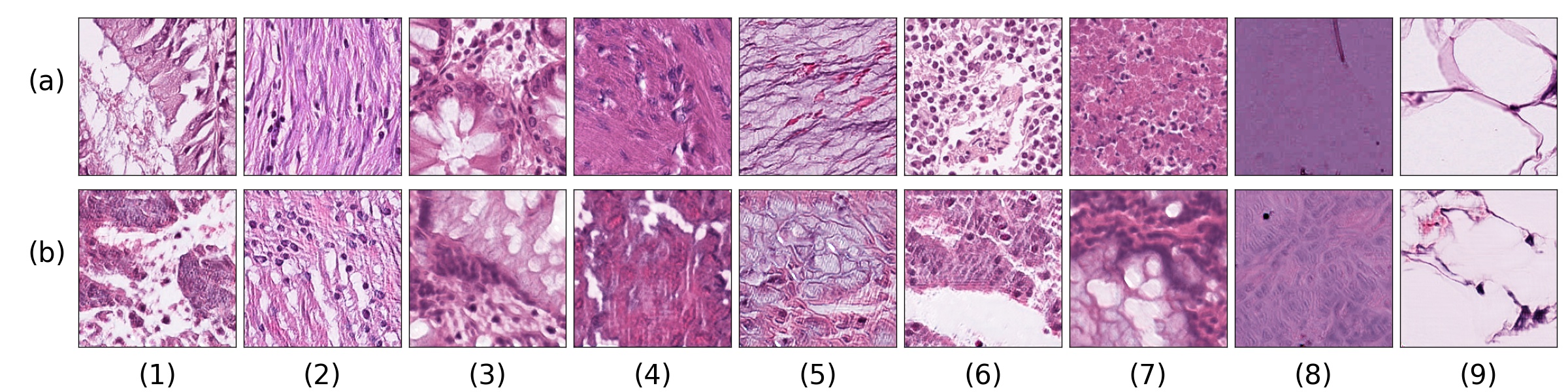}
        \includegraphics[scale=0.24]{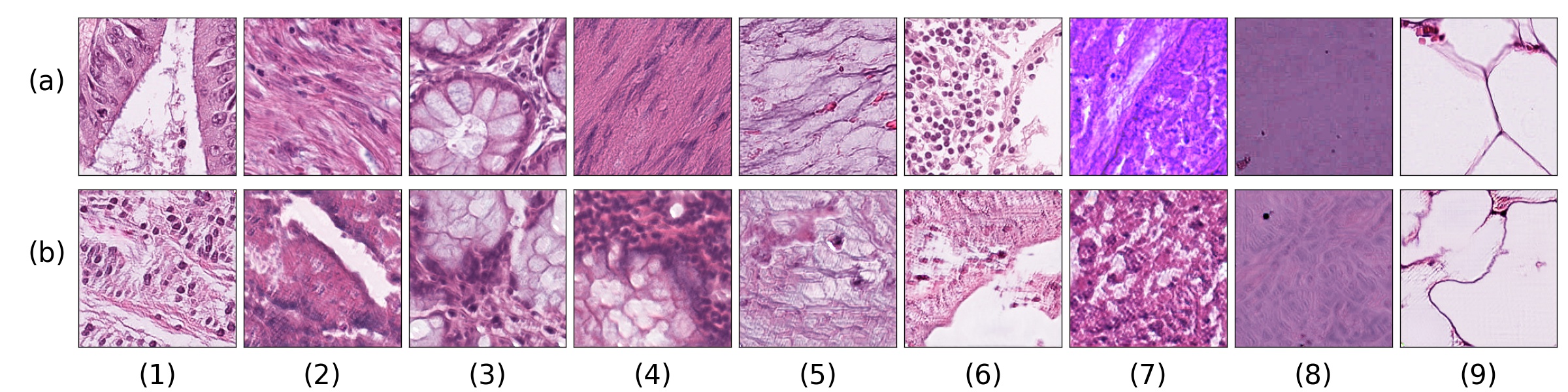}
        \includegraphics[scale=0.24]{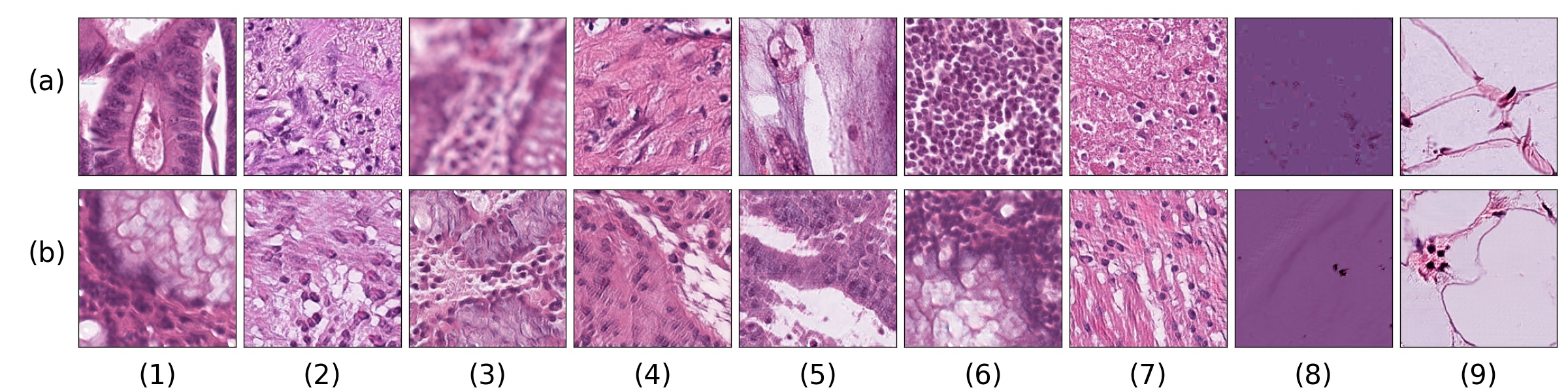}
        \includegraphics[scale=0.24]{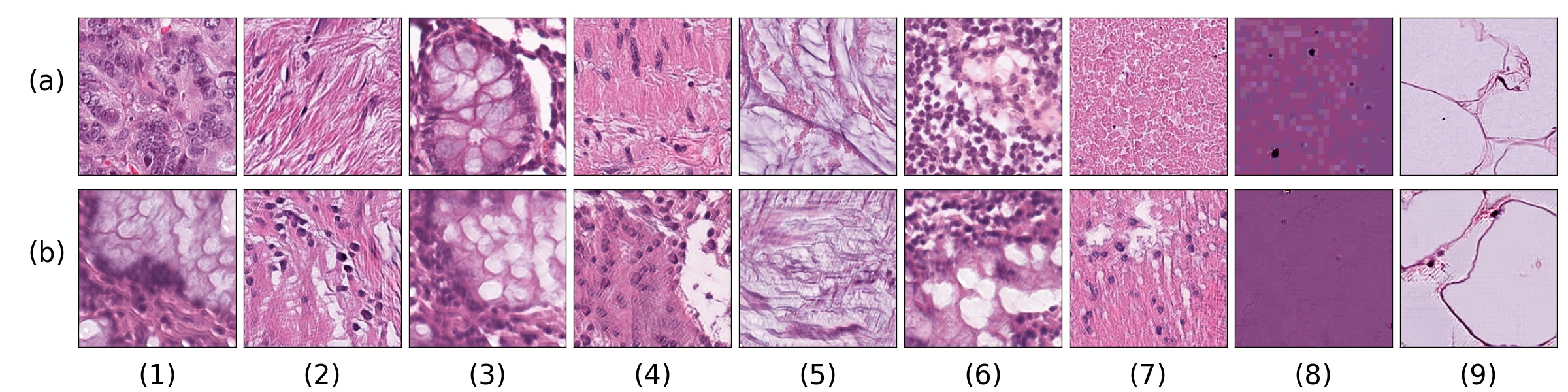}
        \caption{Real tissue images and their reconstructions. We take real tissue images and map them to the latent space with our encoder, then we use the generator with the latent vector representations to generate the image reconstructions. (a) correspond to the real tissue images $X_{real}$ and (b) to the reconstructions $X_{recon}=G(E(X_{real}))$, the images are paired in columns. We present samples of colorectal cancer tissue from NCT with different tissue types: (1) colorectal adenocarcinoma epithelium (tumor), (2) cancer-associated stroma, (3) normal colon mucosa, (4) smooth muscle, (5) mucus, (6) lymphocytes, (7) debris, (8) background, and (9) adipose.}
        \label{fig:appendix_real_recon_nct}
    \end{figure}
	
	\begin{figure}[H]
        \centering
        \includegraphics[scale=0.24]{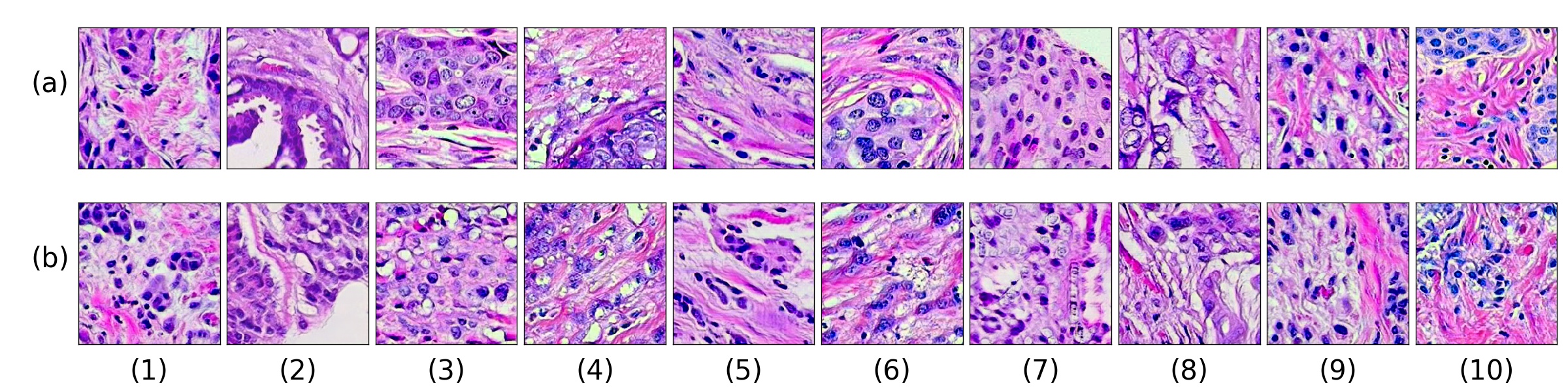}
        \includegraphics[scale=0.24]{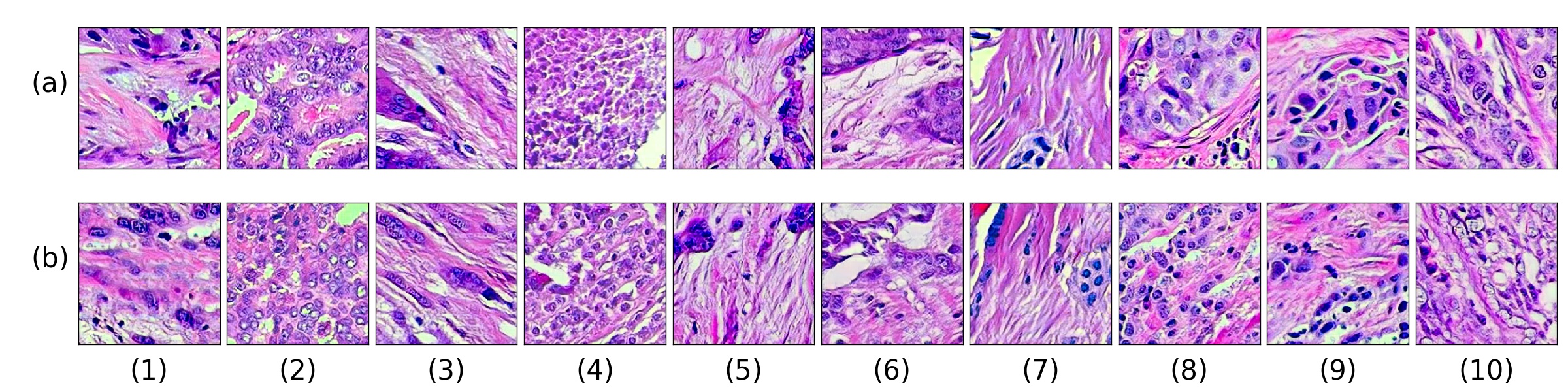}
        \includegraphics[scale=0.24]{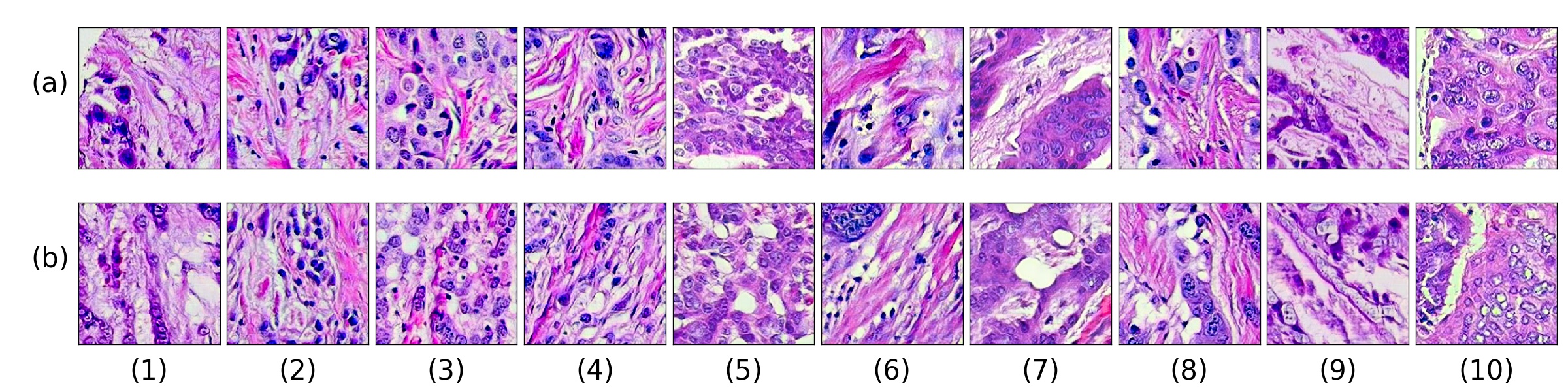}
        \includegraphics[scale=0.24]{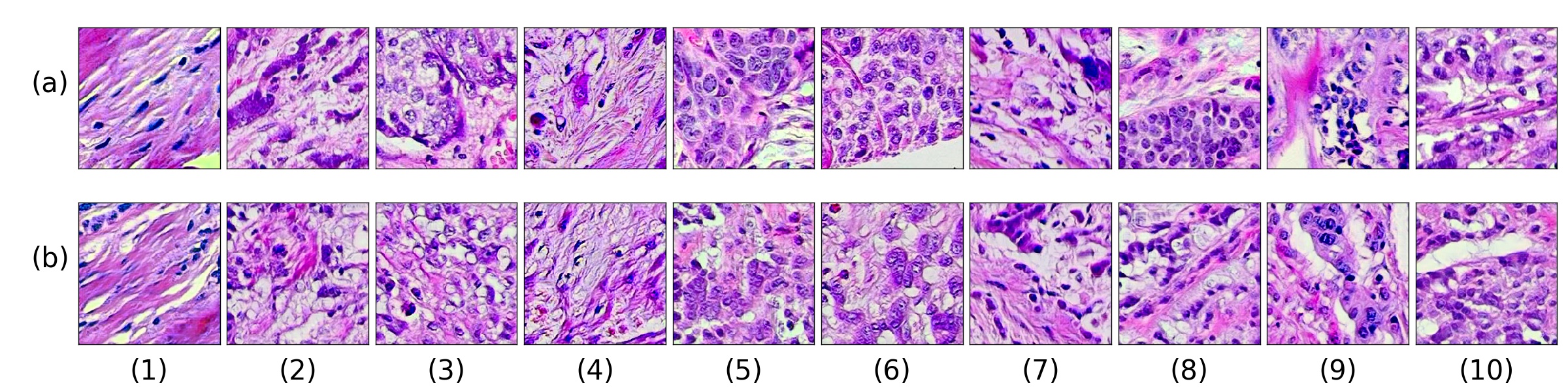}
        \caption{Real tissue images and their reconstructions. We take real tissue images and map them to the latent space with our encoder, then we use the generator with the latent vector representations to generate the image reconstructions. (a) correspond to the real tissue images $X_{real}$ and (b) to the reconstructions $X_{recon}=G(E(X_{real}))$, the images are paired in columns. We present samples of breast cancer tissue from VGH and NKI.}
        \label{fig:appendix_real_recon_vgh_nki}
    \end{figure}

\section{Logistic Regression Confusion Matrix and ROC}
\label{Appendix:roc}
    \begin{figure}[H]
		\centering
		\includegraphics[scale=0.3,clip,trim=0cm 1cm 0cm 3.5cm]{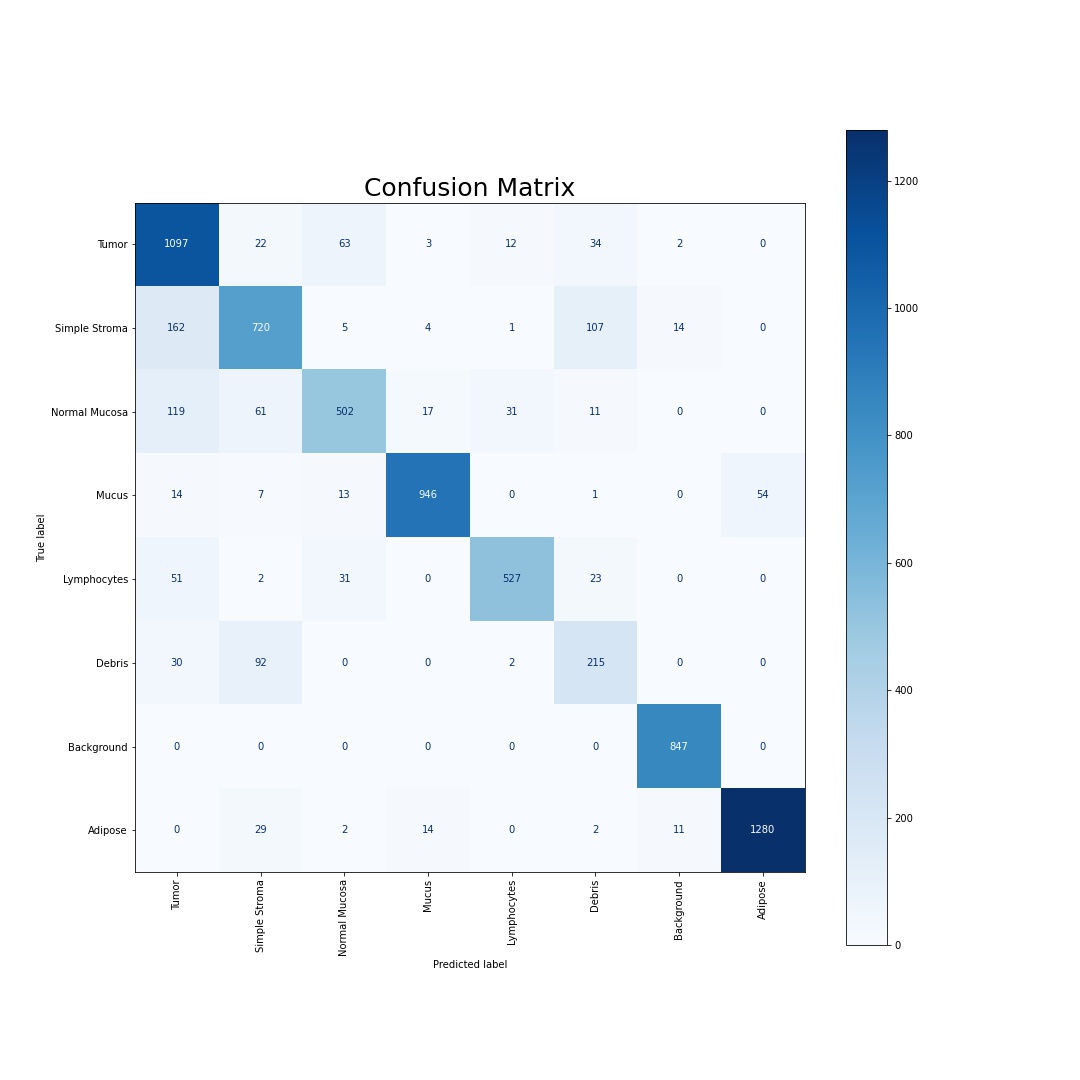}
        \includegraphics[scale=0.35,trim=0cm 2.5cm 0cm 0cm]{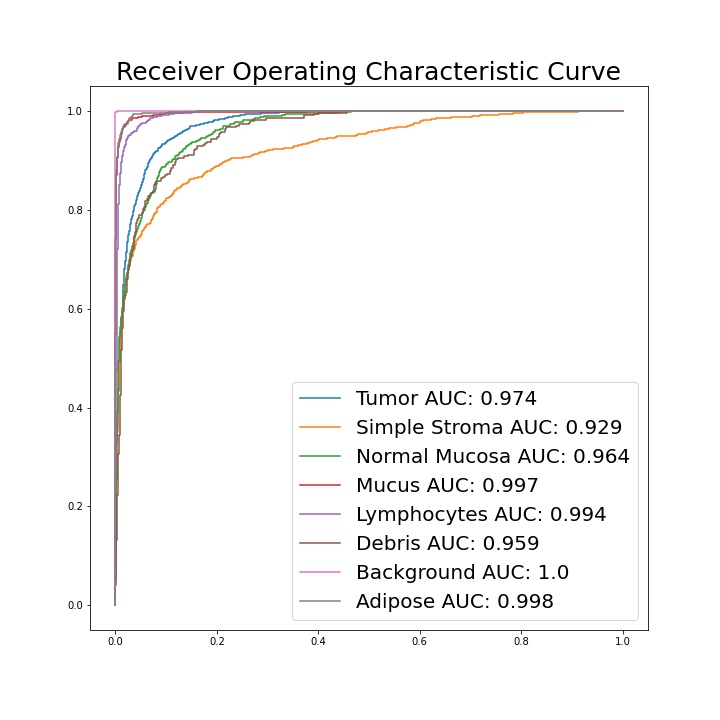}
		\caption{Confusion matrix and Receiver Operating Characteristic (ROC) curve of a logistic regression classifier for tissue type, trained over tissue representations.}
		 \label{fig:roc_confusion_matrix}
	\end{figure}    
    	
\section{Network Architectures}
\label{Appendix:networks}

    \begin{table}[H]
        \centering
        \begin{tabular}{c}
        Generator Network $G:w \rightarrow x$ \\
        \toprule
        \midrule
        Dense Layer, adaptive instance normalization (AdaIN), and leakyReLU \\
        $200 \rightarrow 1024$ \\
        \midrule
        Dense Layer, AdaIN, and leakyReLU \\
        $1024 \rightarrow 12544$ \\
        \midrule
        Reshape $7\times7\times256$ \\
        \midrule
        ResNet Conv2D Layer, 3x3, stride 1, pad same, AdaIN, and leakyReLU $0.2$ \\
        $ 7\times7\times256 \rightarrow 7\times7\times256 $ \\
        \midrule
        ConvTranspose2D Layer, 2x2, stride 2, pad upscale, AdaIN, and leakyReLU $0.2$ \\
        $ 7\times7\times256 \rightarrow 14\times14\times512 $ \\
        \midrule
        ResNet Conv2D Layer, 3x3, stride 1, pad same, AdaIN, and leakyReLU $0.2$ \\
        $ 14\times14\times512 \rightarrow 14\times14\times512 $ \\
        \midrule
        ConvTranspose2D Layer, 2x2, stride 2, pad upscale, AdaIN, and leakyReLU $0.2$ \\
        $ 14\times14\times512 \rightarrow 28\times28\times256 $ \\
        \midrule
        ResNet Conv2D Layer, 3x3, stride 1, pad same, AdaIN, and leakyReLU $0.2$ \\
        $ 28\times28\times256 \rightarrow 28\times28\times256 $ \\
        \midrule
        Attention Layer at $28\times28\times256$ \\
        \midrule
        ConvTranspose2D Layer, 2x2, stride 2, pad upscale, AdaIN, and leakyReLU $0.2$ \\
        $ 28\times28\times256 \rightarrow 56\times56\times128 $ \\
        \midrule
        ResNet Conv2D Layer, 3x3, stride 1, pad same, AdaIN, and leakyReLU $0.2$ \\
        $ 56\times56\times128 \rightarrow 56\times56\times128 $\\
        \midrule
        ConvTranspose2D Layer, 2x2, stride 2, pad upscale, AdaIN, and leakyReLU $0.2$ \\
        $ 56\times56\times128 \rightarrow 112\times112\times64 $ \\
        \midrule
        ResNet Conv2D Layer, 3x3, stride 1, pad same, AdaIN, and leakyReLU $0.2$ \\
        $ 112\times112\times64 \rightarrow 112\times112\times64 $ \\
        \midrule
        ConvTranspose2D Layer, 2x2, stride 2, pad upscale, AdaIN, and leakyReLU $0.2$ \\
        $ 112\times112\times64 \rightarrow 224\times224\times32 $ \\
        \midrule
        Conv2D Layer, 3x3, stride 1, pad same, $ 32 \rightarrow 3 $ \\
        $ 224\times224\times32 \rightarrow 224\times224\times3 $ \\
        \midrule
        Sigmoid \\
        \bottomrule
        \bottomrule
        \end{tabular}
        \vspace*{2mm}
        \caption{Generator Network Architecture details of Pathology GAN model.}
        \label{generator_arch}
    \end{table}
	
	\newpage
    
    \begin{table}[H]
        \centering
        \begin{tabular}{c}
        Discriminator Network $C:x \rightarrow d$ \\
        \toprule
        \midrule
        $x \in  \mathbb{R}^{224\times224\times3}$ \\
        \midrule
        ResNet Conv2D Layer, 3x3, stride 1, pad same, and leakyReLU $0.2$ \\
        $ 224\times224\times3 \rightarrow 224\times224\times3 $ \\
        \midrule
        Conv2D Layer, 2x2, stride 2, pad downscale, and leakyReLU $0.2$ \\
        $ 224\times224\times3 \rightarrow 122\times122\times32 $ \\
        \midrule
        ResNet Conv2D Layer, 3x3, stride 1, pad same, and leakyReLU $0.2$ \\
        $ 122\times122\times32 \rightarrow 122\times122\times32 $ \\
        \midrule
        Conv2D Layer, 2x2, stride 2, pad downscale, and leakyReLU $0.2$ \\
        $ 122\times122\times32 \rightarrow 56\times56\times64 $ \\
        \midrule
        ResNet Conv2D Layer, 3x3, stride 1, pad same, and leakyReLU $0.2$ \\
        $ 56\times56\times64 \rightarrow 56\times56\times64 $ \\
        \midrule
        Conv2D Layer, 2x2, stride 2, pad downscale, and leakyReLU $0.2$ \\
        $ 56\times56\times64 \rightarrow 28\times28\times128 $ \\
        \midrule
        ResNet Conv2D Layer, 3x3, stride 1, pad same, and leakyReLU $0.2$ \\
        $ 28\times28\times128 \rightarrow 28\times28\times128 $ \\
        \midrule
        Attention Layer at $28\times28\times128$ \\
        \midrule
        Conv2D Layer, 2x2, stride 2, pad downscale, and leakyReLU $0.2$ \\
        $ 28\times28\times128 \rightarrow 14\times14\times256 $ \\
        \midrule
        ResNet Conv2D Layer, 3x3, stride 1, pad same, and leakyReLU $0.2$ \\
        $ 14\times14\times256 \rightarrow 14\times14\times256 $ \\
        \midrule
        Conv2D Layer, 2x2, stride 2, pad downscale, and leakyReLU $0.2$ \\
        $ 14\times14\times256 \rightarrow 7\times7\times512 $ \\
        \midrule
        Flatten $ 7\times7\times512 \rightarrow 25088 $ \\
        \midrule
        Dense Layer and leakyReLU, $25088 \rightarrow 1024$ \\
        \midrule
        Dense Layer and leakyReLU, $1024 \rightarrow 1$ \\
        \bottomrule
        \bottomrule
        \end{tabular}
        \vspace*{2mm}
        \caption{Discriminator Network Architecture details of Pathology GAN model.}
        \label{Discriminator_arch}
    \end{table}
	
	\newpage
    
    \begin{table}[H]
        \centering
        \begin{tabular}{c}
        Encoder Network $E:x \rightarrow w'$ \\
        \toprule
        \midrule
        $x \in  \mathbb{R}^{224\times224\times3}$ \\
        \midrule
        Conv2D Layer, 2x2, stride 2, pad downscale, Instance Norm, and leakyReLU $0.2$ \\
        $ 224\times224\times3 \rightarrow 224\times224\times32 $ \\
        \midrule
        ResNet Conv2D Layer, 3x3, stride 1, pad same, Instance Norm, and leakyReLU $0.2$ \\
        $ 224\times224\times32 \rightarrow 224\times224\times32 $ \\
        \midrule
        Conv2D Layer, 2x2, stride 2, pad downscale, Instance Norm, and leakyReLU $0.2$ \\
        $ 224\times224\times32 \rightarrow 122\times122\times64 $ \\
        \midrule
        ResNet Conv2D Layer, 3x3, stride 1, pad same, Instance Norm, and leakyReLU $0.2$ \\
        $ 122\times122\times64 \rightarrow 122\times122\times64 $ \\
        \midrule
        Conv2D Layer, 2x2, stride 2, pad downscale, Instance Norm, and leakyReLU $0.2$ \\
        $ 122\times122\times64 \rightarrow 56\times56\times128 $ \\
        \midrule
        ResNet Conv2D Layer, 3x3, stride 1, pad same, Instance Norm, and leakyReLU $0.2$ \\
        $ 56\times56\times128 \rightarrow 56\times56\times128 $ \\
        \midrule
        Conv2D Layer, 2x2, stride 2, pad downscale, Instance Norm, and leakyReLU $0.2$ \\
        $ 56\times56\times128 \rightarrow 28\times28\times256 $ \\
        \midrule
        ResNet Conv2D Layer, 3x3, stride 1, pad same, Instance Norm, and leakyReLU $0.2$ \\
        $ 28\times28\times256 \rightarrow 28\times28\times256 $ \\
        \midrule
        Attention Layer at $28\times28\times256$ \\
        \midrule
        Conv2D Layer, 2x2, stride 2, pad downscale, Instance Norm, and leakyReLU $0.2$ \\
        $ 28\times28\times256 \rightarrow 14\times14\times512 $ \\
        \midrule
        ResNet Conv2D Layer, 3x3, stride 1, pad same, Instance Norm, and leakyReLU $0.2$ \\
        $ 14\times14\times512 \rightarrow 14\times14\times512 $ \\
        \midrule
        Conv2D Layer, 2x2, stride 2, pad downscale, Instance Norm, and leakyReLU $0.2$ \\
        $ 14\times14\times512 \rightarrow 7\times7\times512 $ \\
        \midrule
        Flatten $ 7\times7\times512 \rightarrow 25088 $ \\
        \midrule
        Dense Layer and leakyReLU, $25088 \rightarrow 1024$ \\
        \midrule
        Dense Layer and leakyReLU, $1024 \rightarrow 200$ \\
        \bottomrule
        \bottomrule
        \end{tabular}
        \vspace*{2mm}
        \caption{Encoder Network Architecture details of Pathology GAN model.}
        \label{Encoder_arch}
    \end{table}
    
    \begin{table}[H]
        \centering
        \begin{tabular}{c}
        Mapping Network $M:z \rightarrow w$ \\
        \toprule
        \midrule
        $z \in  \sim \mathbb{R}^{200} \sim \mathcal{N}(0, I)$ \\
        \midrule
        ResNet Dense Layer and ReLU, $200 \rightarrow 200$ \\
        \midrule
        ResNet Dense Layer and ReLU, $200 \rightarrow 200$ \\
        \midrule
        ResNet Dense Layer and ReLU, $200 \rightarrow 200$ \\
        \midrule
        ResNet Dense Layer and ReLU, $200 \rightarrow 200$ \\
        \midrule
        Dense Layer, $200 \rightarrow 200$ \\
        \bottomrule
        \bottomrule
        \end{tabular}
        \vspace*{2mm}
        \caption{Mapping Network Architecture details of Pathology GAN model.}
        \label{mapping_network_arch}
    \end{table}
	
	\newpage

\end{document}